\documentclass[aps,groupedaddress]{revtex4}

\usepackage{graphics,epsfig}

\def\bit{\begin{itemize}}
\def\eit{\end{itemize}}
\def\ben{\begin{enumerate}}
\def\een{\end{enumerate}}

\def\beq{\begin{equation}}
\def\eeq{\end{equation}}

\def\bea{\begin{eqnarray}}
\def\eea{\end{eqnarray}}

\newenvironment{Eqnarray}%
     {\arraycolsep 0.14em\begin{eqnarray}}{\end{eqnarray}}

\def\beqa{\begin{Eqnarray}}
\def\eeqa{\end{Eqnarray}}
\def\beqno{\begin{eqalignno}}
\def\eeqno{\end{eqalignno}}
\def\ifmath#1{\relax\ifmmode #1\else $#1$\fi}
\def\lsim{\mathrel{\raise.3ex\hbox{$<$\kern-.75em\lower1ex\hbox{$\sim$}}}}
\def\gsim{\mathrel{\raise.3ex\hbox{$>$\kern-.75em\lower1ex\hbox{$\sim$}}}}

\def\anti{\overline}

\def\anti{\bar}

\def\beqno{\begin{eqalignno}}
\def\eeqno{\end{eqalignno}}
\def\beq{\begin{equation}}
\def\eeq{\end{equation}}

\def\ifmath#1{\relax\ifmmode #1\else $#1$\fi}

\def\mw{m_W}

\def\mt{m_t}
\def\mb{m_b}

\def\muw{\mu_W}
\def\mub{\mu_b}

\begin{document}

\title{Leptonic decays of the $B$ charged meson and $B \to X_s \gamma$ in the two Higgs doublet model type III}
\author{J. P. Id\'arraga}
\affiliation{Departament de Physique, Universit\'e de Montr\'eal, Montr\'eal, Canada}
\affiliation{Departamento de F\' isica, Universidad Nacional de Colombia, Bogota, Colombia}
\author{R. Mart\'{\i}nez} 
\affiliation{Departamento de Fisica, Universidad Nacional de Colombia\\
Bogota, Colombia}
\author{N. Poveda}
\affiliation{Escuela de Fisica, Universidad Pedagogica y Tecnologica de Tunja, Tunja, Colombia}
\affiliation{Departamento de Fisica, Universidad Nacional de Colombia\\
Bogota, Colombia}
\author{J-Alexis Rodr\'{\i}guez } 
\affiliation{Departamento de Fisica, Universidad Nacional de Colombia\\
Bogota, Colombia}



\begin{abstract}
We consider the Two Higgs Doublet Model (2HDM) of type III which leads to
Flavour Changing Neutral Currents (FCNC) at tree level. In the framework of this model we calculate the NLO contribution for $b \to s \gamma$ and the branchings for the meson decays $B^+ \to l^+ \nu$. We examine the limits on the new parameters $\lambda_{bb}$ and $M_{H^\pm}$. We take into account the relationship between $\lambda_{tt}$ and $\lambda_{bb}$ coming from the validness of perturbation theory. 
\end{abstract}

\maketitle

The Standard Model (SM) of particle physics based on the gauge group $SU(3)_c \times SU(2)_L \times U(1)_Y$ makes fit
the symmetry breaking by including a fundamental weak doublet of scalar Higgs bosons $\phi$ with a scalar potential  $V(\phi)= \lambda (\phi^\dagger \phi - \frac 12 v^2)^2$. However, the SM does not explain the dynamics responsible for the generation of masses. Between the spectrum of extensions of the SM, many of them include more than one scalar Higgs doublet; for instance, the case of the minimal supersymmetric standard model (MSSM)). We consider a prototype of these extensions of the SM which are including a richer scalar sector,  called generically the Two Higgs Doublet Model (2HDM). There are several kinds of such 2HDM models. In the model called type I, one Higgs doublet provides masses to the up and down quarks, simultaneously. In the model type II, one Higgs doublet gives masses to the up type quarks and the other one to the
down type quarks. These two models include
a discrete symmetry  to avoid flavour changing neutral currents (FCNC) at tree level \cite{gw}. However, the addition of these discrete symmetries is not compulsory and in this case both doublets are contributing to generate the masses for up-type and down-type quarks. In the literature, such a model is known as the 2HDM type III \cite{III}. It has been used to search for physics beyond the SM and specifically for FCNC at tree level \cite{ARS,Sher, we}. In general, both doublets can acquire a vacuum
expectation value (VEV), but  one of them can be absorbed redefining the Higgs boson fields properly. Nevertheless, other studies on 2HDM-III using different basis have been done and there is a case where both doublets get VEVs that allows to study the models type I and II in a specific limit \cite{haber,we}.

In the 2HDM models, the two complex Higgs doublets correspond to eigth scalar states. Spontanoues Symmetry breaking procedure leads to five Higgs fields: two neutral CP-even scalars $h^0$ and $H^0$, a neutral CP-odd scalar $A^0$, and two charged scalars $H^\pm$.
While the neutral Higgs bosons may be difficult to distinguish from the one of the SM, the charged Higgs bosons would have a distinctive signal for physics beyond the SM. Therefore the direct or indirect evidence of a charged Higgs boson would play an important role in the discovery of an extended Higgs sector. Direct searches have carried out by LEP collaborations and they reported a combined lower limit on $M_{H^\pm}$ of 78.6 GeV \cite{lep} assuming $H^+ \to \tau^+ \nu_\tau(c \bar s)$. At the Tevatron, the direct searches for charged Higgs boson are based on $p \bar p \to t \bar t$ where at least one top quark is using the channel $t \to H^+ b$. The CDF collaboration has reported a direct search for charged Higgs boson, setting an upper limit on $B(t \to H^+b)$ around  0.36 at 95 \% C.L. for masses in the range of 60-160 GeV \cite{cdf}. On the other hand, indirect and direct searches have been done by D0 looking for a decrease  in the $t \bar t \to W^+ W^- b \bar b$ signal expected from the SM and also the direct search for the decay mode $H^\pm \to  \tau^\pm \nu$. We should note that all these bounds have been gotten in the framework of the 2HDM type II. And, in the framework of the 2HDM type II and MSSM a full one loop calculation of $\Gamma (t \to b H^+)$ including all sources for large Yukawa couplings were presented in references \cite{qcd,otros}. Other experimental bounds on the charged Higgs boson mass come from processes where the charged Higgs boson is a virtual particle which is the case of the process $b \to s \gamma$. However, the indirect limits which have been obtained from the measurement of the branching ratio $B \to X_s \gamma$ are strongly model dependent \cite{bsg-exp}. Finally, the search for the charged Higgs boson will continue above the top quark mass at LHC. The main production mechanisms would be the processes $gg \to tb H^+$ and $gb \to t H^+$ which have been studied using simulations of the LHC detectors \cite{assamagan}. 

The charged Higgs boson can also be revealed through contributions to low energy processes such as $B^0-\anti{B^0}$, $D^0-\anti{D^0}$ and $K^0-\anti{K^0}$ and bounds on the charged Higgs sector have been found \cite{III}. Moreover, there are other options through leptonic decays of the charged $B$ mesons. They occur via the annihilation process $B^\pm \to W^* (H^*) \to l^\pm \nu_l$. Then, it is possible to use the upper limits on these branching ratios obtained at CLEO \cite{cleo}, BELLE \cite{belle} and BABAR \cite{babar} in order to get bounds on the charged Higgs boson mass. Moreover, recent experimental result on $B(B_u \to \tau \nu)$ were reported by BELLE \cite{belle2} and it is the first evidence of this kind of decays. The decays $B^\pm \to l^\pm \nu_l$ are sensitive at tree level to charged Higgs bosons and can be enhanced up to the current experimental limits \cite{akeroyd,isidori} by multi-Higgs models. On the other hand, the rare decay $B \to X_s \gamma$ is sensitive to  charged higgs bosons at one loop level through electromagnetic and chromomagnetic penguin diagrams, and  therefore the decay $B \to X_s \gamma$ can put strong constraints on the parameters of any charged Higgs sector because its high precision measurement done by CLEO \cite{cleo2}. 

In the present work, we study the processes $B \rightarrow X_s \gamma$  and $B^+ \to l^+ \nu$ in the framework of the 2HDM type III. And we concentrate on the charged Higgs boson sector of this model, with the relevant parameters being its mass $M_{H^\pm}$ and the coupling intensities  $\lambda_{ij}$.

The 2HDM type III is an extension of the SM which adds a new Higgs doublet and three new Yukawa couplings in the quark and leptonic sectors. The mass terms for the up-type or down-type sector depend on two Yukawa coupling matrices. The rotation of the quarks and lepton gauge eigenstates allow us to diagonalize one of the matrices but not both simultaneously, then one of the Yukawa coupling matrix remains non-diagonal, generating the FCNC at tree level. The Higgs couplings to fermions are model dependent. The most general structure for the Higgs-fermion Yukawa couplings in the so called 2HDM type-III  \cite{III} is as follow:
\begin{eqnarray}
-\pounds _{Y} &=&\eta _{ij}^{U,0}\overline{Q}_{iL}^{0}\widetilde{\Phi }%
_{1}U_{jR}^{0}+\eta _{ij}^{D,0}\overline{Q}_{iL}^{0}\Phi _{1}D_{jR}^{0}+\eta
_{ij}^{E,0}\overline{l}_{iL}^{0}\Phi _{1}E_{jR}^{0} \nonumber \label{Yukawa} \\
&+&\xi _{ij}^{U,0}\overline{Q}_{iL}^{0}\widetilde{\Phi }_{2}U_{jR}^{0}+\xi
_{ij}^{D,0}\overline{Q}_{iL}^{0}\Phi _{2}D_{jR}^{0}+\xi _{ij}^{E,0}\overline{%
l}_{iL}^{0}\Phi _{2}E_{jR}^{0}\nonumber \\
&+& h.c. 
\end{eqnarray}
where $\Phi _{1,2}\;$ are the Higgs doublets, $\widetilde{\Phi}_i\equiv
i\sigma_2 \Phi^*_i$,
$Q_L^0 $ is the weak isospin quark doublet, 
and $U^0_R$, $D^0_R$ are weak isospin quark singlets, whereas $\;\eta _{ij}^{0}\;$ and $\xi
_{ij}^{0}\;$ are non-diagonal $3\times 3\;$ non-dimensional matrices and $i$, 
$j$ are family indices. The superscript $0$ indicates that the fields are
not mass eigenstates yet. In the so-called model type I, the discrete
symmetry forbids the terms proportional to $\eta _{ij}^{0},\;$ meanwhile in
the model type II the same symmetry forbids terms proportional to $\xi_{ij}^{D,0},\;\eta _{ij}^{U,0},\xi _{ij}^{E,0}$. We are considering the 2HDM-III in a basis where only one Higgs doublet acquire VEV and then it does not have the usual parameter $\tan \beta=\nu_2/\nu_1$ of the 2HDM type II. In this way we have the usual 2HDM type III \cite{Sher}, where the Lagrangian of the charged sector is given by 
\beq
-L^{III}_{H^\pm ud}=H^+ \anti U [ K \xi^D P_R-\xi^U K P_L] D+h.c.
\eeq
where $K$ is the Cabbibo-Kobayashi-Maskawa (CKM) matrix and $\xi^{U,D}$ the flavour changing matrices. In the framework of the 2HDM type III is useful the parameterization proposed by Cheng and Sher \cite{Sher} for the couplings $\xi_{ii}=\lambda_{ii} g m_i/(2m_W)$.

The leptonic decays of the $B^\pm$ mesons are possible via annihilation processes into $W^\pm$ bosons or $H^\pm$ bosons, the first one is the usual SM contribution and the second one in our case is own to the 2HDM type III. Its amplitude is proportional to the product of the CKM matrix element $V_{ub}$ and the $B$ meson decay constant $f_B$. We should mention that the branching fractions for $e^- \bar \nu_e$ and $\mu \bar \nu_\mu$ in the framework of the SM are helicity suppressed by factors of $\sim 10^{-8}$ and $\sim 10^{-3}$, respectively. But physics beyond the SM can enhance these branching fractions through the introduction of a charged Higgs boson, as we will notice.  The decay width can be written as
\beq
\Gamma(B^\pm \to l^\pm \nu_l)^{III}= \frac{G_F^2 m_B m_l^2 f_B^2}{8 \pi} \vert V_{ub} \vert^2 \left( 1- \frac{m_l^2}{m_B^2} \right) \left[ 1- \frac{\vert d \vert \vert b \vert M_B}{2 \sqrt{2} G_F m_l m^2_H}\right]^2
\label{blnu}
\eeq
where in the framework of the 2HDM-III, we have the factors
\begin{eqnarray}
d &=& \xi_{ll} \\
b &=& \frac{g}{2 m_W } V_{ub} m_b \lambda_{bb}.
\end{eqnarray}
In this form the decay width depends only on the free parameters $\xi_{ll}$, $m_{H^\pm}$ and $\lambda_{bb}$. About the experimental data for the $B$ meson decays $B^- \to l^- \bar \nu_l$, they are experimentally challenging because there are at least two undectetable neutrinos in the final state. These kind of decays has been searched at BELLE, BABAR and CLEO-b.  Bounds on the braching fraction $B(B \to \mu \nu$ have been reported, the stringent bounds come from BABAR measurements and they are \cite{babar} $B(B \to \mu \nu_\mu)\leq 6.8 \times 10^{-6}$ and the SM prediction is $B(B \to \mu \nu_\mu)=3.9 \times 10^{-7}$. About the decay $B_u \to \tau \nu$, the first evidence has been reported by BELLE \cite{belle2}, they report an experimental result of $B(B \to \tau \nu_\tau)= 1.06^{+0.34}_{-0.28}(stat)^{+0.18}_{-0.16}(sys) \times 10^{-4}$. In addition, the values predicted by the SM are  and $B(B \to \tau \nu_\tau)= 1.59 \times 10^{-4}$, which is consistent with the experiment within errors. This new measurement could  guide to a deeper undertanding of flavour and electroweak dynamics, and it could provide evidence of a non-standard Higgs sector. As we already mentioned in the $B$ meson decays is possible to reduce the number of parameters to $\lambda_{bb}$ and the charged Higgs boson mass $m_H$ where we have used the flavour changing couplings for the leptonic sector from the literature\cite{radiaz}. These couplings are bounded by $-0.12 \leq \xi_{22} \leq 0.12$ and  $-1.8 \times 10^{-2} \leq \xi_{33} \leq 2.2 \times 10^{-2}$. Then we can show the plane $\lambda_{bb}$-$m_H$ for the $B$ meson decays under study.  The figure 1 shows the allowed values for $\lambda_{bb}$ vs $m_H$ according to the experimental result from BELLE \cite{belle2} for the $B \to \tau \nu_\tau$ decay, they correspond to the region above curve.  There is another  curve inside this region that  corresponds to the values of the parameters $\lambda_{bb}$ and $m_H$ that can predict the same value of the SM. It is when the factor of equation (\ref{blnu}), $(1-dbM_B/(2\sqrt 2 G_F m_l m_H^2))^2$ is equal to one. These values there, in the same plane,  are indicating that the two models in such conditions are not distinguishable. Something quite similar can be gotten using the experimental bound for $B(B \to \mu \nu)$. 

\begin{figure}[htbp]
\begin{center}
\includegraphics[width=8cm,height=8cm, angle=0]{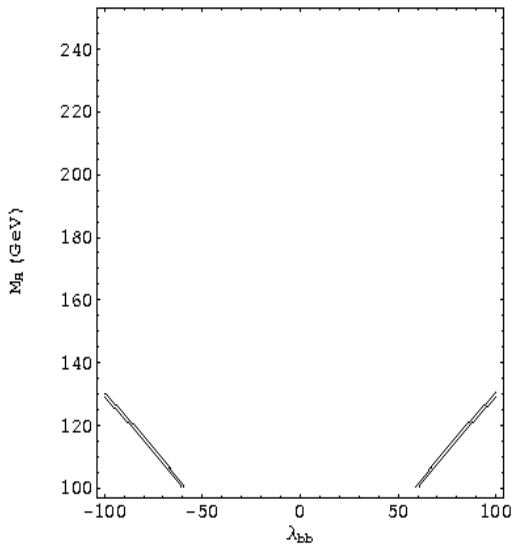}
\end{center}
\caption{The plane $\lambda_{bb}$-$M-H$ for the $B \to \tau \nu_\tau$  decay in the 2HDM-III, it also shows the SM values which are the upper plot.}
\end{figure}

On the other hand, for the radiative decay $B \to X_s \gamma$ we follow references \cite{chinos, borzumati}. The $B \to X_s \gamma$ process as any FCNC process does not arise at the tree level in the SM. In the framework of the SM it is generated by the one-lopp W-exchange diagrams but these contributions are small enough to be comparable to nonstandard contributions, in our case the exchange of a charged scalar Higgs boson. The branching ratio of the inclusive radiative decay $B \to X_s \gamma$ is
\beq
B(B \to X_s \gamma)_{LO}=B_{SL} \vert \frac{V_{ts}^* V_{tb}}{V_{cb}} \vert^2 \frac{6 \alpha_{em}}{\pi f(z)} \vert C_7^{0,eff} (\mu_b)\vert^2
\eeq
at the leading order level, where $C_7^{0,eff}(\mu_b)$ is the effective coefficient at the scale $\mu_b$, 
\beq
C^{0,\,{\rm eff}}_7(\mub) = 
  \eta^\frac{16}{23}  C^{0,\,{\rm eff}}_7(\muw)
 +\frac{8}{3} \left(\eta^\frac{14}{23} -\eta^\frac{16}{23}\right)
                   C^{0,\,{\rm eff}}_8(\muw)
 + \sum_{i=1}^8 h_i \,\eta^{a_i} \,C^{0,\,{\rm eff}}_2(\muw) \,,
\eeq

$f(z)=1-8z^2+8z^6-z^8-24z^4 \log z$ is the phase space factor in the semileptonic  $b$-decay  parameterized in terms of 
$z=m_c^{pole}/m_b^{pole}$ and $\alpha_{em}$ is the fine-structure constant. The coefficients $C_{7,8}^{0,eff}(\mu_b)$ have an important property and it is that they are quite similar in many interesting extensions of the SM, such as 2HDM or the MSSM \cite{grinstein,bertolini,borzumati} and therefore it is possible to parametrize  the new contributions using new fuctions$C_{i,j}^{0,1}(\mu_W)$ with $i=7,8$ and $j=YY, XY$. These functions depend on the unknown  parameter $m_H^{\pm}$ and also on the size and sign fo the couplings $X$ and $Y$ that in the case of the model III under study they are $X=-\lambda_{bb}$, and $Y=\lambda_{tt}$. To get these couplings we assume that the flavour changing parameters for the light quarks are negligible and $\lambda_{bb} > 1$, $\lambda_{tt}<1$ which is the case disccussed by Atwood, Reina and Soni as their third case \cite{ARS}. Then the LO Wilson coefficients at the matching energy scale $\mw$ are \cite{chinos, borzumati},
\bea
 C^{0,\,{\rm eff}}_2(\muw)  & = &  1\, ,                        \nonumber \\
 C^{0,\,{\rm eff}}_i(\muw)  & = &  0,  \hspace*{1cm} (i=1,3,4,5,6)\, , \nonumber \\
 C^{0,\,{\rm eff}}_7(\muw)  & = &   C_{7,SM}^0(\mw)   + |Y|^2  \, C_{7,YY}^0(\mw)
  +        (XY^*) \, C_{7,XY}^0(\mw) \, , \label{eq:c70mw}  \nonumber \\
 C^{0,\,{\rm eff}}_8(\muw)  & = &   C_{8,SM}^0(\mw )  + |Y|^2  \, C_{8,YY}^0(\mw)
  +        (XY^*) \, C_{8,XY}^0(\mw) \, ,
\eea
with
\bea
 C_{7,SM}^0  & = & \frac{3x_{t}^3-2x_{t}^2}{4(x_{t}-1)^4}\ln{x_t}
                     +\frac{-8x_{t}^3-5x_{t}^2+7x_t}{24(x_t-1)^3},
  \nonumber   \\
 C_{8,SM}^0  & = &\frac{-3x_{t}^2}{4(x_{t}-1)^4}\ln{x_t}
                     +\frac{-x_{t}^3+5x_{t}^2+2x_t}{8(x_t-1)^3}\, ,
 \nonumber \\
 C_{7,YY}^0  & = & \frac{3y_{t}^3-2y_{t}^2}{12(y_{t}-1)^4}\ln{y_t}
                     +\frac{-8y_{t}^3-5y_{t}^2+7y_t}{72(y_t-1)^3}\, ,
 ,\nonumber \\
 C_{7,XY}^0 & = & \frac{y_t}{12} \left[
 \frac{-5y_t^2+8y_t-3+(6y_t-4)\ln y_t}{(y_t-1)^3} \right]\, ,\label{eq:c70-xy}\\
 C_{8,YY}^0  & = & \frac{-3y_{t}^2}{12(y_{t}-1)^4}\ln{y_t}
                     +\frac{-y_{t}^3+5y_{t}^2+2y_t}{24(y_t-1)^3} \, ,\nonumber \\
 C_{8,XY}^0 & = & \frac{y_t}{4}  \left[ \frac{-y_t^2+4y_t-3- 2 \ln y_t}{(y_t-1)^3} \right] \,,
\eea
where $x_t=m_t^2/\mw^2$, $y_t=m_t^2/m_h^2$, and these leading order functions have no explicit $\muw$ dependence.  

Now, at the next leading order level that is neccesary in order to use the experimental data, the branching ratio is
\beq  
B(B \to X_s \gamma)_{NLO}=B_{SL} \vert \frac{V_{ts}^* V_{tb}}{V_{cb}} \vert^2 \frac{6 \alpha_{em}}{\pi f(z) \kappa(z)} [\vert \bar D \vert^2+A+\Delta]
\label{nlo}
\eeq
where $B_{SL}$ is the measured semileptonic branching ratio of B mesons, and  $\kappa(z)$ is the QCD correction for the semileptonic $B$ decay. The term $\bar D$ corresponds to the subprocesses $b \to s \gamma$ which involves the NLO Wilson coefficient $C_7^{eff}(\mu_b)$, the virtual correction functions $r_i$ and $\gamma_{i7}^{0,eff}$ the elements of the anomalous dimension matrix which govern the evolution of the Wilson coefficents from the matching scale $\mu_W$ to lower scale $\mu_b$. The term $A$ in equation (\ref{nlo}) is the correction coming from the bremsstrahlung process $b \to s \gamma g$ and in the $\Delta$ have been included the nonperturbative corrections obtained  with the method of the heavy-quark effective theory relating the actual hadronic process to the quark decay rate.  The whole set of functions already mentioned have been given in references \cite{chinos, borzumati}. With the set of above equations we can estimate the ratio $B(B \to X_s \gamma)$ and use the experimental world average $B(B \to X_s \gamma)^{exp}=(3.52 \pm 0.30) \times 01^{-4}$ \cite{worldBS}.   

\begin{figure}[htbp]
\begin{center}
\includegraphics[width=10cm, height=8cm,angle=0]{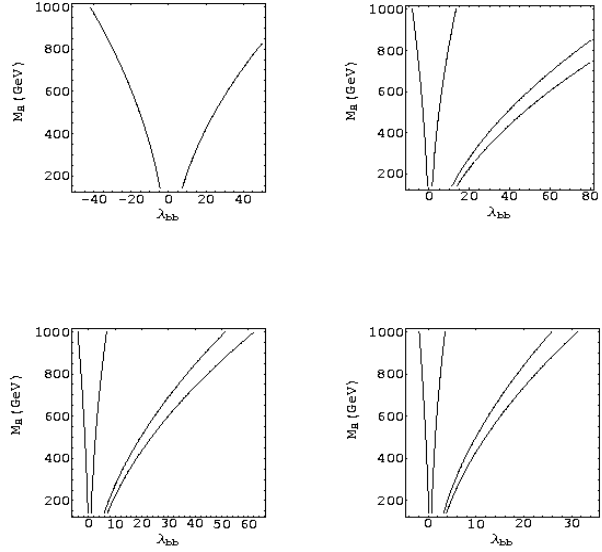}
\end{center}
\caption{The allowed values in the plane $\lambda_{bb}$-$M_H$ taking into account the experimental values for $B(B_s \to X_s \gamma$ and  $B(B \to \tau \nu)$. Different values of $\lambda_{tt}=(0.1,0.5,1,2)$.} 
\end{figure}

\begin{figure}[htbp]
\begin{center}
\includegraphics[width=10cm, height=8cm, angle=0]{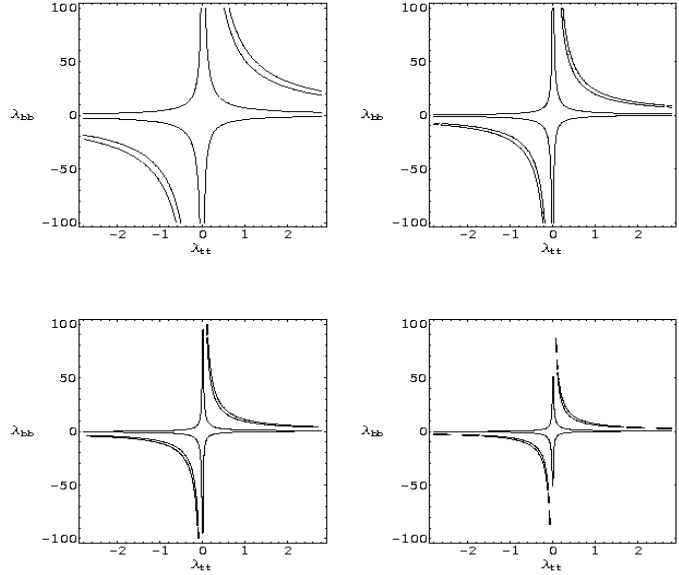}
\end{center}
\caption{The  plane $\lambda_{bb}$-$\lambda_{tt}$ taking into account the experimental value for $B(B_s \to X_s \gamma$ and $B(B \to \tau \nu)$. Using different values of the charged Higgs boson mass $M_H=(120,250,500,1000)$ GeV}
\end{figure}

In figure 2, we present the allowed regions in the plane $M_H$ versus $\lambda_{bb}$ for different values of $\lambda_{tt}$ which is appearing in the $B \to X_s \gamma$ decay at NLO order. And in figure 3, we present the allowed regions for the $B(B \to X_s \gamma)$ in the plane $\lambda_{bb}$-$\lambda_{tt}$ for different values of the charged Higgs boson mass. 
But in order to get  with the numerical evaluations, we are going to take into account the possible values of $\lambda_{bb,tt}$ which are consistent with perturbation theory. And from the perturbation theory considerations we have already gotten the inequality \cite{rozo}
\beq
\frac{\mb^2}{\mt^2} \vert \lambda_{bb} \vert^2 + \vert \lambda_{tt} \vert^2 <8 ,\label{uno}
\eeq 
where we have used the equation (2) and the parameterization  proposed in reference \cite{Sher} for the couplings $\xi_{ii}$. It defines an ellipse with $\vert \lambda_{bb}\vert \leq 100$ and $\vert \lambda_{tt} \vert \leq \sqrt{8}$. In this case we consider the inequality from perturbation theory validness in order to reduce the space of parameters, equation (\ref{uno}).  This link between the parameter $\lambda_{tt}$ and $\lambda_{bb}$ allows to get the plane $\lambda_{bb}$ versus $m_H$ using the experimental measurement for the branching ratio $B(B_s \to X_s \gamma )$.  Finally, in figure 4 we show the case of the induced decay $B \to X_s \gamma$ decay.  The fullfilled regions are the allowed regions, it means these are the regions satisfying the experimental region and the perturbation theory constraint. We notice that these regions in figure 4 correspond to a different choice of $\lambda_{tt}$ as it was presented in figure 2.

\begin{figure}[htbp]
\begin{center}
\includegraphics[angle=0]{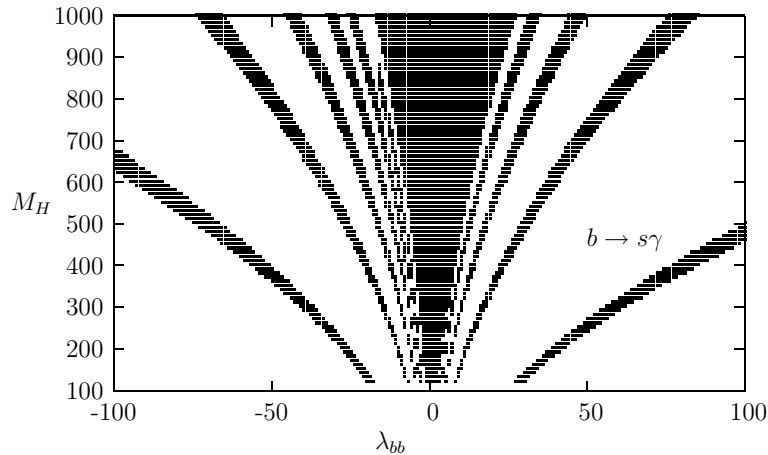}
\end{center}
\caption{The allowed values in the plane $\lambda_{bb}$-$m_H$ taking into account the experimental value for $B(B_s \to X_s \gamma$ and the condition coming from validness of perturbation theory.}
\end{figure}

We have studied in the framework of the 2HDM type III, the allowed region for the parameters $\lambda_{bb}$ and $m_H$ using the processes $B \to \tau \nu_\tau$, $B  \to \mu \nu_\mu$ and $B \to X_s \gamma$. During the study, we have used the condition on the parameter space coming from the fact that the Yukawa couplings should be perturbative, equation (\ref{uno}), in to order to reduce the number of free parameters. Finally, we  have compared the plots looking  for the stringest regions in the plane $\lambda_{bb}$-$m_H$ and we have noticed that the $B \to X_s \gamma$ decay is the most restrictive process constraining the parameters of the charged Higgs sector in the 2HDM-III. But however there are small regions for small values of $\lambda_{bb}$ and light $m_H$ that leptonic decays can exclude. We also have found that in case of the leptonic decays $B \to l \nu$,  there are values of the parameters $\lambda_{bb}$ and $m_H$ given a 2HDM predition which cannot be distinguishable from the SM prediction. It is because the factor $(1 - d b M_B/ (2 \sqrt 2 G_F m_l m_H^2))^2$ in equation(\ref{blnu})could get the value equal one for some values of $\lambda_{bb}$ and $m_H$ and then reach out the SM prediction. Therefore, these values must be in the allowed experimental region of the plane $\lambda_{bb}-m_H$ as it was showed in figure 1.  
    
We acknowledge to Carlos Sandoval for useful discussions. This work has been supported by COLCIENCIAS and HELEN.



\begin{thebibliography}{000}

\bibitem{gw}  S. Glashow and S. Weinberg, Phys. Rev. D \textbf{15}, 1958
(1977).

\bibitem{III}  W.S. Hou, Phys. Lett B \textbf{296}, 179 (1992); D. Cahng, W.
S. Hou and W. Y. Keung, Phys. Rev. D \textbf{48}, 217 (1993); S. Nie and M. Sher, Phys. Rev. D \textbf{58}, 097701 (1998); M. Sher and Y. Yuan, Phys. Rev. D \textbf{44}, 1461 (1991); D. Atwood, L. Reina and A. Soni, Phys. Rev. D \textbf{55}, 3156 (1997); Phys. Rev. Lett. \textbf{75}, 3800 (1995).

\bibitem{ARS}  D. Atwood, L. Reina and A. Soni, Phys. Rev. D \textbf{53},
1199 (1996); Phys. Rev. D \textbf{54}, 3296 (1996); Phys. Rev. Lett. 
\textbf{75}, 3800 (1993); D. Atwood, L. Reina and A. Soni, Phys. Rev. D \textbf{55}, 3156 (1997); G. Cvetic, S. S. Hwang and C. S. Kim, Phys. Rev. D \textbf{58}, 116003 (1998).

\bibitem{Sher}  Marc Sher and Yao Yuan, Phys. Rev. D \textbf{44}, 1461
(1991); T.P. Cheng and M. Sher, Phys. Rev. D \textbf{35}, 3490
(1987)

\bibitem{we} Rodolfo A. Diaz, R. Martinez and J.-Alexis Rodriguez, Phys. Rev. D \textbf{64}, 033004 (2001); Phys. Rev. D \textbf{63},  095500 (2001).
\bibitem{haber}S.~Davidson and H.~E.~Haber,
  Phys.\ Rev.\ D {\bf 72}, 035004 (2005)
  [arXiv:hep-ph/0504050].
\bibitem{lep} LEP Collaborations, arXiv:hep-ex/0107031.
\bibitem{cdf} CDF Collaboration, F. Abe,\,\textit{et. al}, Phys. Rev. Lett. \textbf{79}, 357 (1997); CDF Collaboration, T. Affolder, \textit{et. al}, Phys. Rev. D \textbf{62}, 012004 (2000).
\bibitem{d0} D0 Collaboration, B. Abbot, \textit{et. al}, Phys. Rev. Lett. \textbf{82}, 4975 (1999); V. Abazov, \textit{et. al}, Phys. Rev. Lett. 88, 151803 (2002).
\bibitem{qcd} A. Mendez and A. Pomarol, Phys. Lett. B \textbf{360}, 47 (1995); C. Li and R. J. Oakes, Phys. Rev. D \textbf{43}, 855 (1991); A. Djouadi and P. Gambino, Phys. Rev. D \textbf{51}, 218 (1995). 
\bibitem{otros}M. Carena, D. Garcia, U. Nierste, C. Wagner, arXiv:hep-ph/9912516; M. Carena, J. Conway, H. Haber and J. Hobbs, arXiv:hep-ph/0010338. 
\bibitem{bsg-exp}W.~S.~Hou,
  Phys.\ Rev.\ D {\bf 48}, 2342 (1993).
\bibitem{cleo}S.~Chen {\it et al.}  [CLEO Collaboration],
  Phys.\ Rev.\ Lett.\  {\bf 87}, 251807 (2001)
  [arXiv:hep-ex/0108032].
\bibitem{assamagan}K.~A.~Assamagan, Y.~Coadou and A.~Deandrea,
  Eur.\ Phys.\ J.\ directC {\bf 4}, 9 (2002)
  [arXiv:hep-ph/0203121].K.~A.~Assamagan and N.~Gollub,
  Eur.\ Phys.\ J.\ C {\bf 39S2}, 25 (2005)
  [arXiv:hep-ph/0406013].
\bibitem{belle}K.~Abe {\it et al.}  [Belle Collaboration],
KEK-PREPRINT-2001-88
{\it Prepared for 20th International Symposium on Lepton and Photon Interactions at High Energies (LP 01), Rome, Italy, 23-28 Jul 2001}

\bibitem{babar}B.~Aubert {\it et al.}  [BABAR Collaboration],
  Phys.\ Rev.\ Lett.\  {\bf 95}, 041804 (2005)
  [arXiv:hep-ex/0407038].
\bibitem{belle2} K. Ikado, {\it et.al}[BELLE Collaboration],hep-ex/0604018.

\bibitem{akeroyd}A.~G.~Akeroyd and S.~Recksiegel,
  J.\ Phys.\ G {\bf 29}, 2311 (2003)
  [arXiv:hep-ph/0306037].

\bibitem{isidori} G. Isidori and P. Paradisi, hep-ph/0605012.

\bibitem{cleo2}
  M.~Artuso {\it et al.}  [Cleo Collaboration],
  Phys.\ Rev.\ Lett.\  {\bf 75}, 785 (1995).
\bibitem{grinstein}B.~Grinstein and M.~B.~Wise,
  Phys.\ Lett.\ B {\bf 201}, 274 (1988); P.~L.~Cho and B.~Grinstein,
  Nucl.\ Phys.\ B {\bf 365}, 279 (1991)
  [Erratum-ibid.\ B {\bf 427}, 697 (1994)].B.~Grinstein, R.~P.~Springer and M.~B.~Wise,
  Phys.\ Lett.\ B {\bf 202}, 138 (1988).
\bibitem{bertolini}S.~Bertolini, F.~Borzumati, A.~Masiero and G.~Ridolfi,
  Nucl.\ Phys.\ B {\bf 353}, 591 (1991);  S.~Bertolini and F.~Vissani,
  Z.\ Phys.\ C {\bf 67}, 513 (1995)
  [arXiv:hep-ph/9403397].
\bibitem{borzumati}F.~M.~Borzumati and C.~Greub,
  Phys.\ Rev.\ D {\bf 58}, 074004 (1998)
  [arXiv:hep-ph/9802391].
\bibitem{chinos}Z.~j.~Xiao and L.~Guo,
  Phys.\ Rev.\ D {\bf 69}, 014002 (2004)
  [arXiv:hep-ph/0309103].
\bibitem{rozo}R.~Martinez, J.~A.~Rodriguez and M.~Rozo,
  Phys.\ Rev.\ D {\bf 68}, 035001 (2003)
  [arXiv:hep-ph/0212236].
\bibitem{radiaz} R. A. Diaz, R. Martinez and C. E. 
Sandoval, [arXiv:hep-ph/0311201] (2003). R. A. Diaz, R. Martinez 
and C. E. Sandoval, [arXiv:hep-ph/0406265] (2004). R. A. Diaz, R. 
Martinez and J-A. Rodriguez, [arXiv:hep-ph/0103050] (2001); R. A. Diaz, R. Martinez and 
J-A. Rodriguez, \textit{Phys. Rev. }\textbf{D63}, 095007 (2001), 
[arXiv:hep-ph/0010149]. 
\bibitem{worldBS} E. Barberio, {\it et. al} [The Heavy Flavor Averaging Group], hep-ex/0603003.

\end{thebibliography}
\end{document}